\documentclass[Afour,sageh,times]{sagej}

\usepackage{moreverb,url}

\usepackage[colorlinks,bookmarksopen,bookmarksnumbered,citecolor=red,urlcolor=red]{hyperref}

\newcommand\BibTeX{{\rmfamily B\kern-.05em \textsc{i\kern-.025em b}\kern-.08em
T\kern-.1667em\lower.7ex\hbox{E}\kern-.125emX}}

\definecolor{darkmagenta}{rgb}{0.55, 0.0, 0.55}
\definecolor{ericcolor}{rgb}{0.0, 0.55, 0.55}

\newcommand{\cg}[1]{CG-#1}
\newcommand{\eg}[1]{DG-#1}
\newcommand{\tech}[1]{duplication#1}

\usepackage{tikz}

\newcommand*\annotatedFigureBoxCustom[8]{\draw[#5,thick,rounded corners] (#1) rectangle (#2);\node at (#4) [fill=#6,thick,shape=circle,draw=#7,inner sep=2pt,font=\sffamily,text=#8] {\textbf{#3}};}
\newcommand*\annotatedFigureBox[4]{\annotatedFigureBoxCustom{#1}{#2}{#3}{#4}{white}{white}{black}{black}}

\newenvironment {annotatedFigure}[1]{\centering\begin{tikzpicture}
    \node[anchor=south west,inner sep=0] (image) at (0,0) { #1};\begin{scope}[x={(image.south east)},y={(image.north west)}]}{\end{scope}\end{tikzpicture}}

\def\volumeyear{2016}

\begin{document}

\runninghead{Smith and Wittkopf}

\title{An Empirical Study on the Relationship Between the Number of Coordinated Views and Visual Analysis}

\author{Juyoung Oh\affilnum{1}, Chunggi Lee\affilnum{1}, Hwiyeon Kim\affilnum{1}, Kihwan Kim\affilnum{1}, Osang Kwon\affilnum{1}, Eric D. Ragan\affilnum{2}, Bum Chul Kwon\affilnum{3}, Sungahn Ko\affilnum{1}}

\affiliation{\affilnum{1}UNIST, Korea}
\affiliation{\affilnum{1}UNIST, Korea\\
\affilnum{2}University of Florida, USA\\
\affilnum{3}IBM Research, USA}

\begin{abstract}
Coordinated Multiple views (CMVs) are a visualization technique that simultaneously presents multiple visualizations in separate but linked views. 
There are many studies that report the advantages (e.g., usefulness for finding hidden relationships) and disadvantages (e.g., cognitive load) of CMVs. 
But little empirical work exists on the impact of the number of views on visual anlaysis results and processes, which results in uncertainty in the relationship between the view number and visual anlaysis.
In this work, we aim at investigating the relationship between the number of coordinated views and users’ analytic processes and results. 
To achieve the goal, we implemented a CMV tool for visual anlaysis. 
We also provided visualization duplication in the tool to help users easily create a desired number of visualization views on-the-fly. 
We conducted a between-subject study with 44 participants, where we asked participants to solve five analytic problems using the visual tool.
Through quantitative and qualitative analysis, we discovered the positive correlation between the number of views and analytic results. 
We also found that visualization duplication encourages users to create more views and to take various analysis strategies.
Based on the results, we provide implications and limitations of our study.
\end{abstract}


\maketitle

\section{Introduction}

Coordinated Multiple views (CMVs)~\cite{North00, Baldonado00} are a visualization technique that simultaneously presents multiple visualizations in separate but linked views. 
CMVs allow users to explore different perspectives of data, to collect and organize facts, to identify hidden relationships and to obtain insights through a variety of interactions coordinated between multiple views~\cite{Baldonado00, North00, Roberts04}. 
In addition, CMVs enable users to take multi-path visual exploration during sensemaking~\cite{Pirolli05} and knowledge generation processes~\cite{Sacha14} by supporting quick-and-easy hypothesis testing and validation processes.

Previous studies have shown the advantages and disadvantages of using CMVs for data analyses.
Several researchers have discussed the effects of using CMVs on users' cognitive loads~\cite{Plumlee06, Ryu03, Convertino03, Jun13}. Plumlee and Ware report that using CMVs increases efficiency during a comparison task~\cite{Plumlee06}. 
On the other hand, convertino et al.~\cite{Convertino03} and Jun et al.~\cite{Jun13} find that users could encounter context-switching and cognitive costs while using CMVs. 
Due to the mixed results, several researchers have attempted to provide practical advice related to view coordination models~\cite{Boukhelifa03}, highlighting strategies~\cite{Griffin15}, and design guidelines~\cite{Baldonado00} for CMVs.
These studies have shown that there are potential benefits but also costs associated with using CMVs.

Despite the findings of previous studies, rare work exists on the impact of the number of views on visual anlaysis. 
Thus, it is not clear if visual anlaysis performance (i.e., analysis speed and result accuracy) is related to the view number, when many views are available for users. 
Few studies concern the effects of the number of views during visual anlaysis~\cite{Jun13, Convertino03, Ryu03}, but they use a fixed number of views (up to four views) in their experiments.
We think it is still difficult to determine whether it is actually more or less effective to generate as many views as possible.

In this work we aim at investigating the relationship between the number of views and visual analysis.
To achieve the goal, we implemented a visual analysis tool with visualization duplication for facilitating `easy' view creation.
Our duplication method allows users create a clone of a visualization view (i.e., child view) from any other visualization views (i.e., parent view) with one click.
A clone view instantiated by \tech{} has the same information as its parent view, such as visualization type, states (i.e., parameters, data, color schemes). %
As such, by utilizing \tech{}, users can easily create a desired number of views on-the-fly without having tedious interactions for purposes of parameter selection~\cite{Roberts14, Saraiya05}. 
We conducted a between-subject study, where we asked 44 participants to perform multidimensional data analysis with the tool. 
In the study, each participant was assigned to one of the following two groups: 1) Duplication Group (DG) which had access to visualization duplication to clone previously created views; and 2) Control Group (CG), which did not have access to duplication.
The experimental results indicated that the users who used the \tech{} 1) created more views, 2) solved problems more accurately, and 3) utilized various problem-solving strategies. 
In addition, the DG users showed interesting annotation usage patterns and more frequently utilized the pin interaction, compared to the CG, which did not have access to duplication; however, no significant results were observed in terms of time spent or users' confidence levels.

\section{Related Work}
\label{sec_relatedwork}
Side-by-side comparisons of problem-solving processes are known to be more accurate than back-and-forth scrolling or modifying within one view. %
Coordinated Multiple Views (CMVs) designs present information to users in several visualization views, where view may use the same or different visualization representation. 
A system that performs side-by-side comparisons across two views is sometimes called a dual-view system~\cite{Roberts07}. 
There are numerous ways to use two views, including overview and detail, focus and context, and small-multiple displays. 
Visualization systems with CMVs facilitate participants' exploration of data through two or more views. 
The key benefits of CMVs include an improvement in participant's task performance, the discovery of unforeseen relationships within given data, and a unification of the desktop~\cite{North00}. 
In general, each view is connected to another for better side-by-side navigation and can be filtered via brushing~\cite{Roberts07}.

Though CMVs have been incorporated into several visualization applications~\cite{Roberts07, North00}, few studies have directly reported the positive effects of CMVs via quantitative measures such as task performance (i.e., reduced task completion time and increased accuracy). 
For instance, Ryu et al.~\cite{Ryu03} and Convertino et al.~\cite{Convertino03} asked participants to answer four questions in their experiment using a pair of visualizations among a parallel coordinates plot (PCP), a scatterplot, and a geographical map. 
The experimental results showed that CMVs helped participants perform search tasks with the pair of PCP and scatterplot in terms of time and accuracy. 
Plumlee and Ware~\cite{Plumlee06} also examined the performance of CMVs compared to a zooming technique, and concluded that additional views are more effective if the data are complex and require more storage space than the capacity of a human's working memory. 
Jun et al.~\cite{Jun13} conducted an experiment that allowed participants to use four sequential or simultaneous views for either monitoring or comparing tasks. 
They found simultaneous multiple views to be more effective than sequential multiple views in terms of both completion time and accuracy, though these studies restricted the total number of views to five in total.
In our experiment, we did not impose the limit on the view number to estimate the relationship between view numbers and analytic results, and place greater emphasis on learning about analysis processes and strategies.

We also consider what types of visualization representations are commonly used in view instances for CMVs.
Arguably, a scatterplot and PCP pair is one of the most popular pairs used for a visual analysis.
Scatterplots encode the data of two variables as points in a 2D Cartesian space (i.e., pairs of ($x_{i}$, $y_{i}$) for i = 1, 2, 3...). 
Then additional attributes are represented by different colors, shapes, sizes, and orientations of the points to better support an efficient analysis. 
Often, scatterplots are extended to a scatterplot matrix (SPLOM)~\cite{Cleveland88} for a multivariate correlation analysis. 
While scatterplots use only two axes (i.e., vertical and horizontal axes), PCPs place multiple vertically axes (in which the attribute values of an item are mapped to a location in each axis and connected) to form polylines. 
Incorporating several advanced visualization techniques (e.g., edge bundling~\cite{Palmas14}), makes PCPs more effective. 
Johansson et al. discuss the evaluation, categorization, and guidelines for future research on PCPs, while Heinrich and Weiskopf present a state-of-the-art report on PCPs~\cite{Heinrich13}. 
We include these visualizations in this work because 1) novice participants can quickly learn how to use them without a visualization background, 2) they are domain-independent, and 3) simple and few interactions can provide insights to participants.

There are many human-subject studies that have investigated not only the analysis outcomes but also the processes by studying analytic footage of participants.
While the final outcomes generated from a visual anlaysis are of great value, prior work has demonstrated the anlaysis processes themselves are also important~\cite{Pike09}. 
Investigating participants' analytic processes used to derive insights through provenance can reveal participants' knowledge generation processes, such as how a given tool helps participants to obtain insights. 
This type of research, which aims to understand participants' reasoning process by visualizing and analyzing participant interactions is referred to as insight provenance research (e.g.,~\cite{Gotz09, Guo16, Ragan16}).
Insight provenance has been applied to many domains, including bioinformatics~\cite{Saraiya05}, financial data~\cite{Jeong08}, intelligence analysis~\cite{ragan2015evaluating,andrews2010space} investigative anlayses~\cite{Kang09, Kwon12}, and hotel selection~\cite{Nguyen16}. 
Ragan et al.~\cite{Ragan16} suggested the importance of clarifying the types and purposes of provenance information studied in visualization research; in this work, we study analytic provenance focusing on users' history of interactions and insights for the purpose of supporting  recall and awareness of state during analysis. 
To study provenance in our research, we capture participant interaction logs.
In particular, to derive high-level strategies as well as low-level interaction patterns, we capture the types (what), the amounts (how many), and the methods (how) of participants’ insight gaining processes through annotations and interaction logs by closely following the insight measurement metrics proposed by Guo et al.~\cite{Guo16} and the interaction taxonomy by Yi et al.~\cite{Yi07}. 
We discuss our detailed methods in Section~\nameref{sec_coding}.

\begin{figure}[t]
    \begin{center}
	\includegraphics[width=.8\columnwidth]{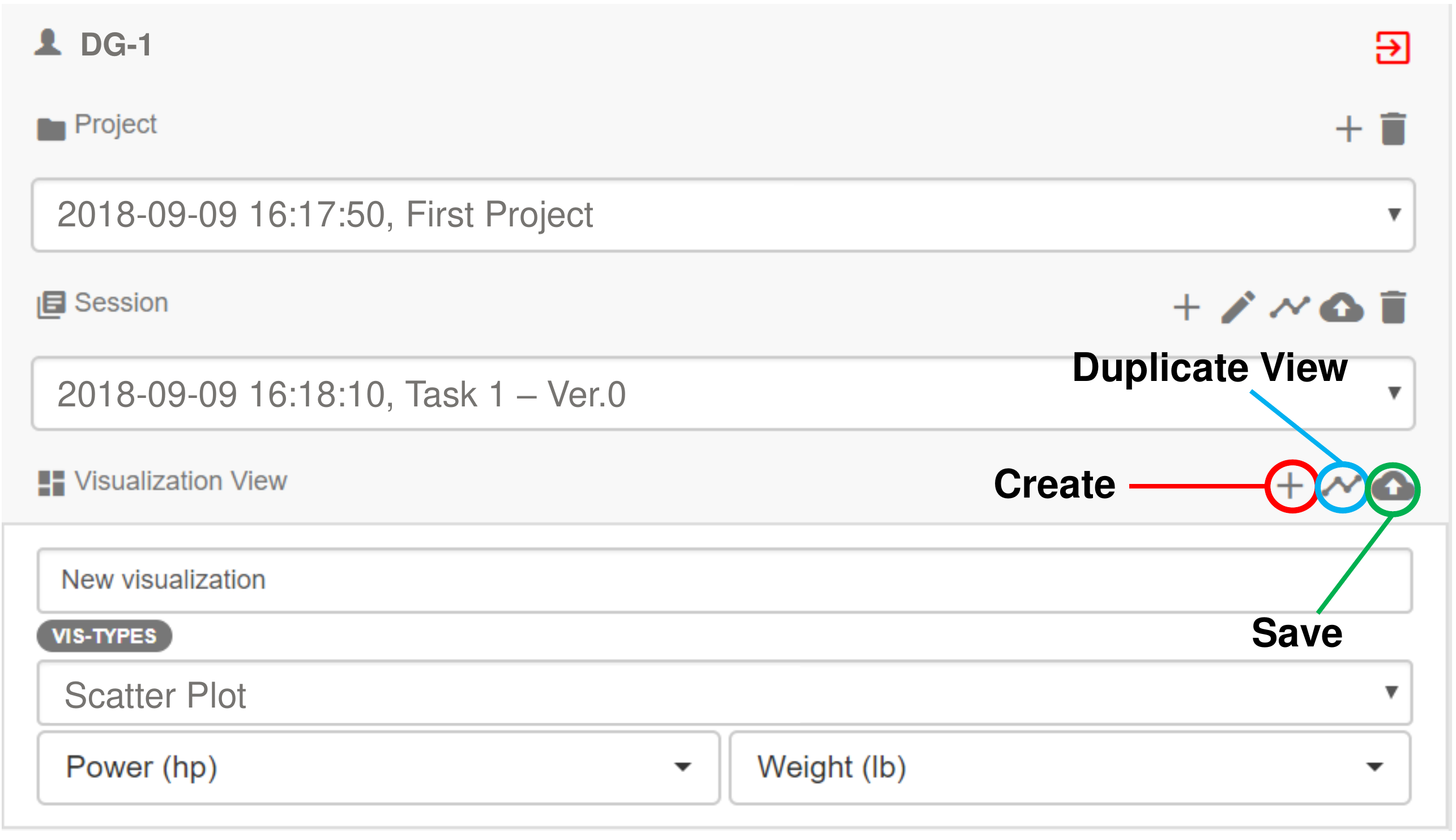}
	\caption{The creation view in the visual tool for our experiment. ``Project" was used in our experiment to distinguish participants--each participant worked on a project that consists of multiple sessions. participants were asked to create ``a session," wherever a new task starts. To create a view, participants needed to 1) click the "+" symbol, 2) select parameters (e.g., view name, visualization type), and ``Save." When the \tech{} button is clicked, participants could create a duplicated view of an existing view.
	}
	\label{fig_view_creation_mode}
	\end{center}
\end{figure}

\section{Visual Tool Design}
\label{sec:tool-design}

In this section, we describe a multidimensional visual analysis tool used for the experiment. 
Two versions of the tool were needed for the two  experimental conditions (with and without the duplication functionality)

\subsection{Overview}
\begin{figure*}[h!t]
\begin{annotatedFigure}
	{\includegraphics[width=1.0\linewidth]{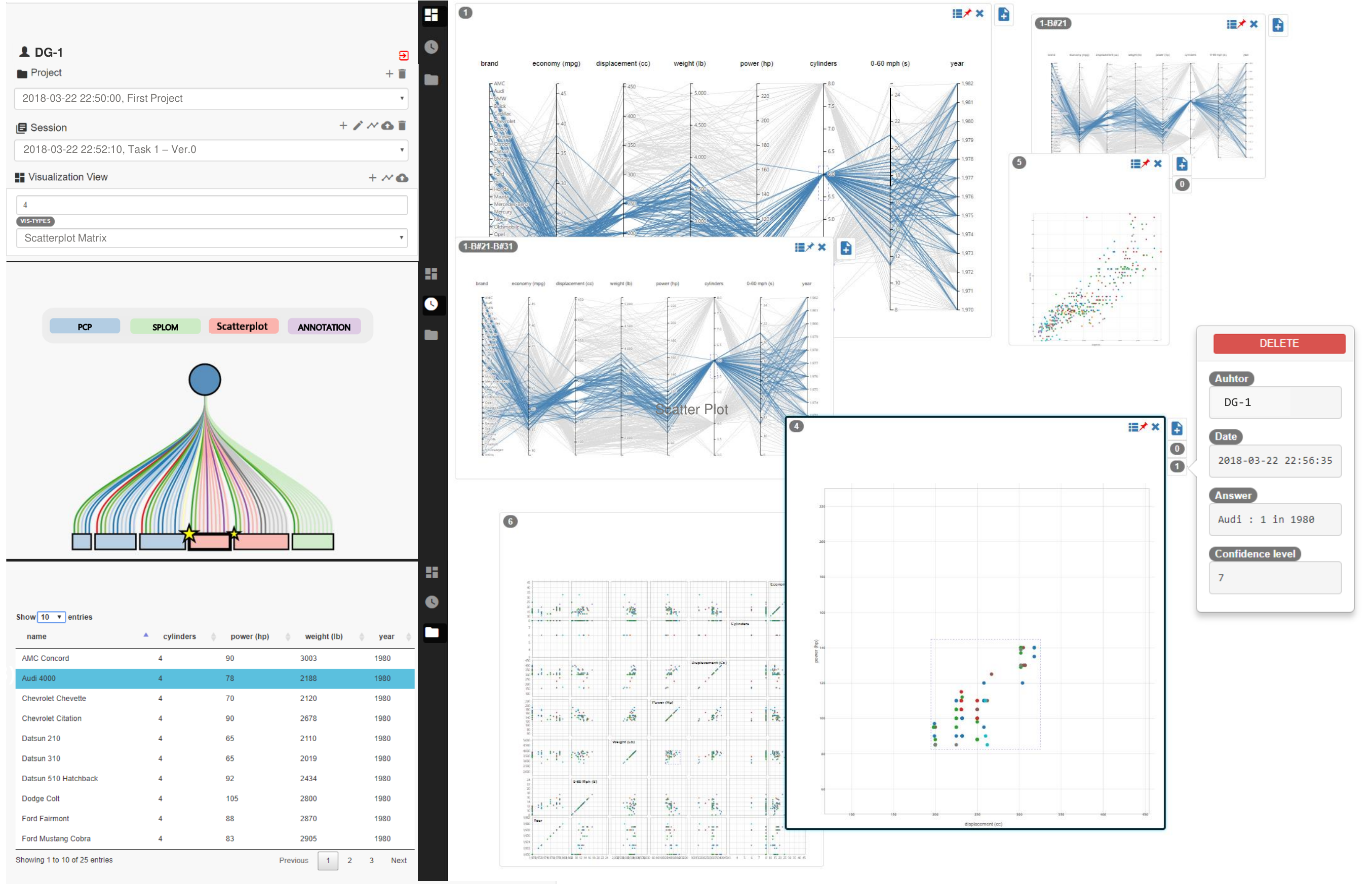}}
	\annotatedFigureBox{0.01,0.6973}{0.301,0.9877}{A}{0.01,0.9877}%
	\annotatedFigureBox{0.009,0.3641}{0.301,0.6959}{B}{0.009,0.6959}%
	\annotatedFigureBox{0.332,0.0001}{1.002,0.9878}{D}{0.332,0.9878}%
	\annotatedFigureBox{0.009,0.0001}{0.301,0.3621}{C}{0.009,0.3621}%
	\annotatedFigureBox{0.868,0.3015}{0.983,0.6244}{E}{0.868,0.6244}%
\end{annotatedFigure}
	\caption{An overview of the prototype tool: (A) View Creation for allowing participants to choose parameters (e.g., visualization types), (B) Visual Work History for representing view and interaction information, (C) Table for viewing filtered or brushed data items, (D) Workspace for problem-solving with created views, and (E) An example annotation template}
	\label{fig_system_overview}
\end{figure*}

As Figure~\ref{fig_system_overview} presents, the visual tool divides the screen into two vertical panels. 
The left-side panel is dedicated to view creation and management: (A) View Creation; (B) Visual Work History; and (C) Table.
On the right side is a larger visualization workspace (D) used for interaction and exploration of data using any  visualization created by the users.
In this panel, participants could also rearrange and make annotations on visualization views.
The tool records all actions performed on each view by logging the action name (e.g., `create a new view by default method'), the view (e.g., `view creation'), and the timestamp.

\subsection{Creation View}

In the Creation View (Figure~\ref{fig_view_creation_mode}), participants are able to instantiate a new view by specifying parameters, namely visualization type and attributes, and to add a title to the visualization.
For instance, a participant can create a scatterplot (type) with weight and power (attributes) with the title of ``Correlation between Weight and Power.''
This method is the default view creation method, and it was available to participants in both study conditions.

Participants in the duplication group had an additional view-duplication method available.
Instead of manually selecting parameters, participants in this group could choose to duplicate a selected view and simply make a copy.
In addition, the newly duplicated view inherits states (e.g., filters set on the view), annotations, and histories (e.g., interactions previously performed on the view).
In the tool, duplication is triggered when a participant clicks a target view and uses the appropriate duplication button (\includegraphics[width=0.04\linewidth]{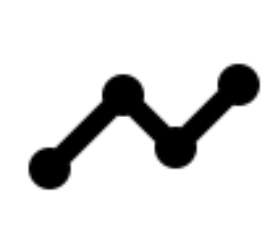}) in the creation view (Figure~\ref{fig_view_creation_mode}). 
When ``Save" (\includegraphics[width=0.04\linewidth]{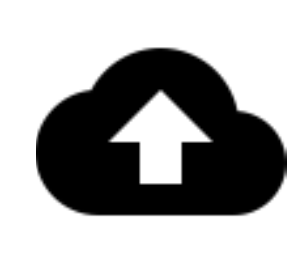}) is selected, the current visual states, parameters, and interaction history of the selected view are saved (i.e., a snapshot of a visualization view was taken) and presented in the workspace.

\begin{figure}[t]
  \centering
  \includegraphics[width=0.95\linewidth]{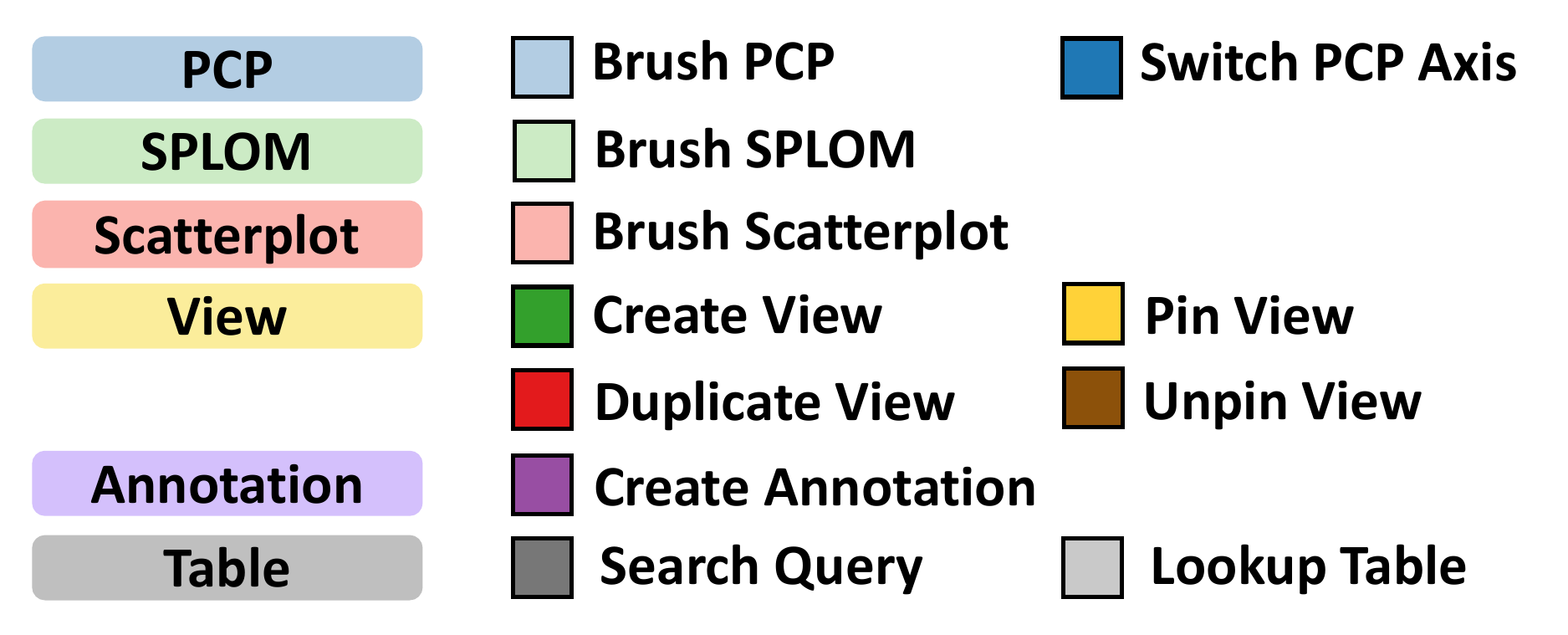}
  \caption{The color mapping for an action taxonomy. %
  }
  \label{fig_action_taxonomy}
\end{figure}

\subsection{Visual Work History (VWH)}
\label{sec:vwh}

Inspired by previous workflow visualizations~\cite{Groth06, Shrinivasan08, Heer08, Dunne12, Maguire12}, we implemented VWH in a tree-shaped visualization to help users monitor their past analysis history, as shown in Figure~\ref{fig_system_overview}--B. 
VWH consists of one parent node at the top and leaves toward the bottom.
Here, leaves indicate visualization views, and links between a parent-child pair are actions performed on the views.
The links and leaves are color-coded according to their corresponding views and actions, as Figure~\ref{fig_action_taxonomy} shows.
This view reflects the history of view creation and actions performed on the views, and it is created in a following manner.
At the beginning, the view only contains the parent node at the top, the blue circle in Fig~\ref{fig_system_overview}--B.
When a participant creates a new view (e.g., parallel coordinates), it creates a link and leaf, representing the `create view' action and the `parallel coordinates (PCP),' respectively.
When the participant performs more actions (e.g., brushes) on the same view, new links are added to the right of the existing links and the leaf increases its width proportionally.
If the participant creates a new view (whether it's from the default method or VD), a new leaf is added to the right and a new link, indicating `creation view' or `duplicate view,' is added to connect the parent node and the new leaf.
When the participant makes an annotation, the view adds a star symbol on the top-left corner of the leaf, and the size of the star is proportional to the number of annotations.
Participants in both study conditions had this view available to help visually track views being created and interactions performed.

\subsection{Table View}

The table view (Figure~\ref{fig_system_overview}-C) shows the items that are selected or filtered in the visualization views.
This view is held separate from the visualization workspace for following reasons.
First, we expected that participants want to regularly inspect actual data details when they perform actions on them.
Second, adjusting the view along with other visualizations is  cumbersome because participants often want the view to be fixed in one location without occlusion.
Lastly, the table view is rarely duplicated or created throughout pilot sessions.
Therefore, we allowed participants to view the table view at a consistent space, and set it to always show selected items so that they could always easily access details for data they were working with the visualization views.

\subsection{Visualization Workspace}

In the Visualization Workspace view (Figure~\ref{fig_system_overview}-D), participants work with views created from the Creation View.
In this view, participants could arrange views freely, resize a view, and bring views to the top layer if occluded.
Within each view created on the Workspace view, participants were able to perform actions like setting filters, sorting items, and switching axes depending on the available actions for corresponding views.
All views being created are coordinated and supported \textit{brushing and linking} such that items highlighted or filtered in one view are also highlighted or filtered in other views. 
Participants could choose to `pin' a view, which freezes the current state of the view regardless of actions performed on other views.
Participants could also unpin any pinned view to make it coordinated with other views again.
In each view, participants could make annotations in order to make personal notes or to answer questions by providing written answers on the text area and their confidence level (7-level Likert scale, 7 being the most confident) about their answer, as Figure~\ref{fig_system_overview}-(E) shows.
In each view, participants could request to check details of selected items in the table view either via brushing or filtering.

Our tool provides three visualizations, Parallel Coordinates (PCPs), Scatterplots, and Scatterplot matrices (SPLOMs) that participants can choose to create and to work with in the visualization workspace.
We chose the techniques that are adequate to solve the given multidmensional data analysis tasks, which will be explained in Section~\nameref{sec:task_description}.
Participants could choose as many attributes (minimum two) as they want for PCPs.
PCPs allow participants to perform two kinds of actions, namely to switch axes, and to set filters by drawing a rectangular region of interest per each axis.
For scatterplots, participants are requested to choose two attributes for two axes at a time.
Scatterplots allow participants to set a filter by drawing a rectangular region of interest on the canvas.
Scatterplot matrices (SPLOMs) show all of the attributes and allow participants to set a filter (scatterplot) by drawing a rectangular region on a cell, which automatically filters out the selected items from all the cells.

\section{Research Questions and Hypotheses}

The goal of this study is to investigate the relationship between the number of views and users' analysis processes and the results obtained.
To facilitate easy view creation, we provided the visualization duplication function. 
Thus, our first research question is whether users create many views by using  \tech{} during their analysis (RQ1).  
Once we confirm RQ1, our next question is how the number of views created by using \tech{} during the analysis affects the accuracy of the analysis results, time spent for anlaysis, user confidence levels, and the number of insights found during the analysis (RQ2). 
Our last question is whether users who created many views with \tech{} show interesting interaction patterns and analysis strategies (RQ3).

With these questions in mind, we derive seven hypotheses to answer the research questions:

\begin{itemize} %
\item [H1] Participants who have access to \tech{} create more views than those who do not (RQ1). %
\item [H2] Participants who have access to \tech{} solve data analysis tasks with higher accuracy than those who do not (RQ2). %
\item [H3] Participants who have access to \tech{} take less time to solve data analysis tasks than those who do not (RQ2). %
\item [H4] Participants who have access to \tech{} generate more insights than those who do not (RQ2). %
\item [H5] Participants who have access to \tech{} are more confident in their analysis processes and results than those who do not (RQ2). %
\item [H6] Participants who have access to \tech{} perform more interactions than those who do not (RQ3). %
\item [H7] Participants who have access to \tech{} use the pin interaction more than those who do not (RQ3). %
\end{itemize}

In the next section, we discuss the experimental design for testing the hypotheses.

\section{Experiment}
\label{sec_experiment}
In this section, we describe the experiment, which compared data analysis between two conditions (with and without view duplication), which were varied following a between-subjects design. 
Participants used the analysis tool (see Section~\nameref{sec:tool-design}) to perform multiple analysis tasks on a multidimensional dataset so we could address our hypotheses.

\subsection{Dataset and Analysis Tasks}
\label{sec:task_description}
We first describe how we selected the data for the experiment. 
First, the experiment required a multidimensional dataset so that participants are inclined to use multiple views. 
Second, due to the available participant pool, the dataset could not require expertise in specific areas for a layperson to understand the context. 
Third, tasks in the experiment should have various levels of difficulty ranging from simple to compound~\cite{Amar04, Valiati06, Kobsa01}. 
Though explicit requirements on the data size are not imposed, a proper size is expected so that participants could obtain meaningful insights and could solve analytic tasks within a given time. 
We decided to use the car dataset~\cite{Grinstein02}, which fulfills the requirements.

As a next step, we collected analytic tasks used in recent studies (2011--2017) with the selected car dataset (e.g.,~\cite{Heinrich12, Kim16, Kuang12, Kwon16, Lee16, Palmas14, Shao17, Walker13, Wall17}). 
By merging the tasks based on task taxonomies from prior literature~\cite{Amar05a, Sarikaya18}, we derived the final task set, which includes five tasks of a mixture of task types.
The five tasks are presented in Table~\ref{table_questions} with their corresponding task types.
Each of the five tasks consist of more than or equal to two task types and encouraged participants to use multiple views. 
Moreover, we designed more advanced and complex tasks (tasks T4 and T5) so that participants were encouraged to save and use multiple analytic paths and to annotate intermediate results along with the multiple views. 
Task T1 asked participants to find anomalies with respect to the relationship between two attributes: power and weight. 
T2 asked participants to compute a ratio from two attributes, power to weight, and to rank cars based on the derived measure. 
T3 asked participants to find the car with the maximum value in one attribute, power, among cars manufactured between 1976 and 1979 by German manufacturers. 
T4 asked participants to find a year that included the greatest number of car models produced by Japanese manufacturers. 
T5 asked participants to find the number of car models produced by German manufacturers in the year when Japanese manufacturers produced the largest variety of car models (the answer to T4). 

At this stage, we repeatedly checked the tasks to determine whether they could be solved using several approaches. 
For example, T3 in Table~\ref{table_questions} can be solved with a PCP in a view by moving two axes and filtering by the power. 
Alternatively, two views can be utilized with one PCP view for brushing and another scatterplot view to find an answer. 
T4 could not be easily solved with one view because it would require a high cognitive load to remember the filtered results and counts. 
Thus, we expected participants to actively use multiple views and annotations.

\begin{table}[t]
	\begin{center}
	\includegraphics[width=0.99\columnwidth]{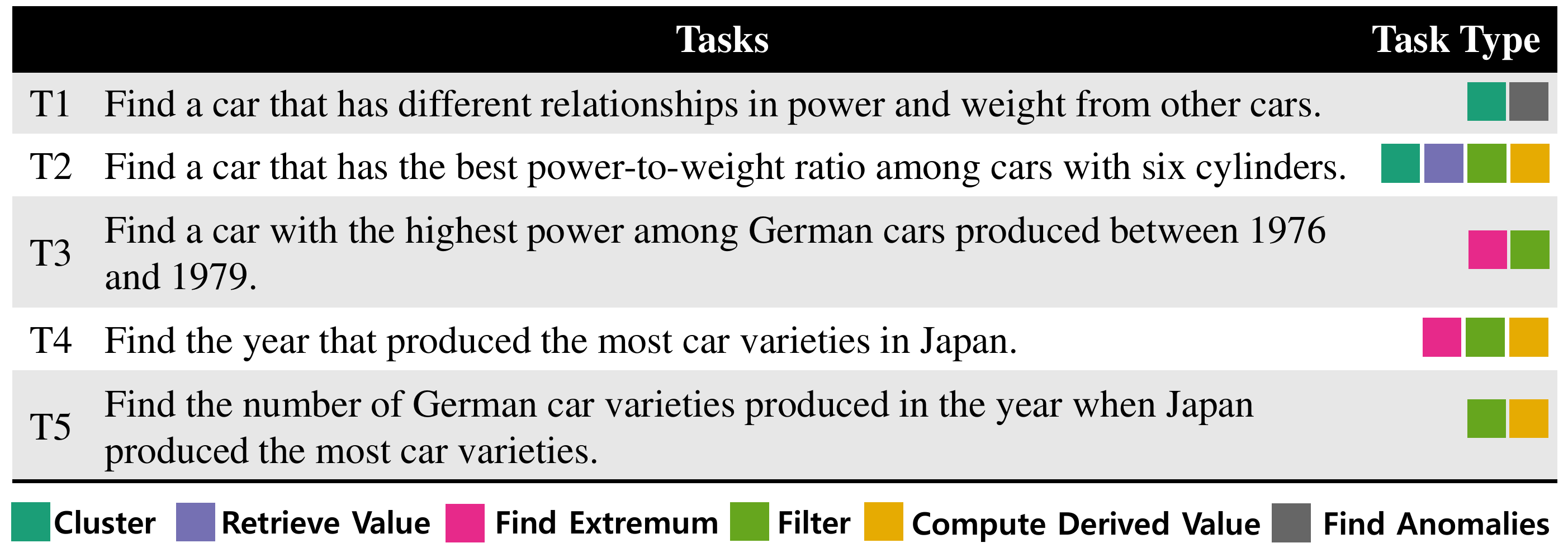}
	 \caption{We create five tasks after a literature survey and task type analysis.}
	\label{table_questions}
	\end{center}
\end{table}

\begin{table*}[t]
    \begin{center}
	\includegraphics[width=2.0\columnwidth]{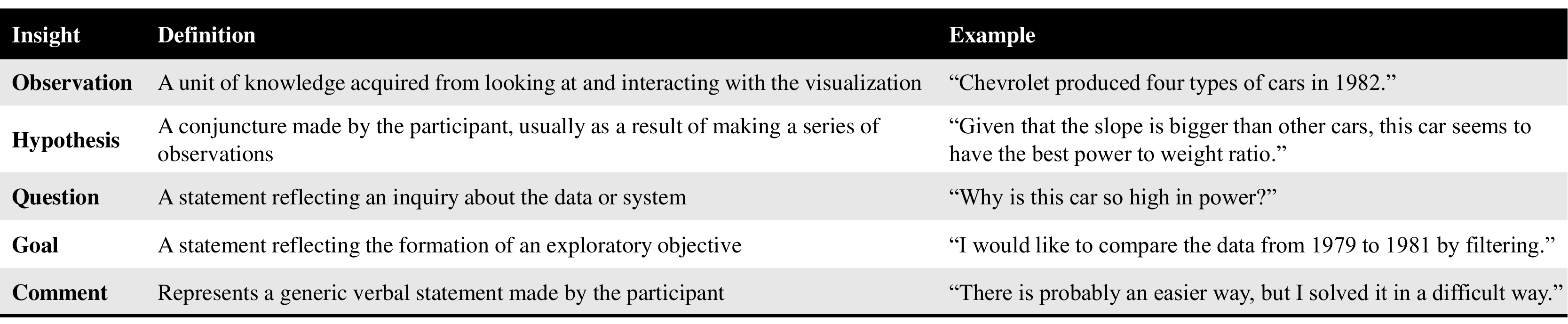}
	\caption{Insight categories, definitions, and examples. We use these five categories of insights for grading the walkthroughs.}
	\label{table_coding_category}
	\end{center}
\end{table*}

\subsection{Participants}

We recruited 44 participants at our university (19 female participants) who were students from science and engineering schools (age range: 19--27).
Participants self-reported an average of 3.83 visualization familiarity score based on a 7-point scale (1 = the least proficient; 7 = the most proficient). 
No participants reported any experience with the car data set.

Next, each participant was assigned to one of two groups: CG (Control Group) or DG (Duplication Group). 
As we expected there could be participants who would choose not to use \tech{}, we increased the number of participants assigned to the DG.
Among the 29 participants in the DG, we observed that 5 participants did not use \tech{} while solving all tasks. 
In the end, we decided to include all participants regardless of using \tech{} or not, because they had the freedom to choose which options to use based on their best judgment and preferences. 
Consequently, we ultimately had a sample of 44 participants for analysis, with 15 participants in the CG and 29 participants in the DG.

\subsection{Procedure}

Before starting the study, all participants first reviewed and signed an agreement form to participate, and then they filled out a form for collecting demographic information (e.g., gender, age, education level).
Next, they watched a brief tutorial video before an experimenter explained the functions of the tool. 
Then they were given time to freely use the tool. 
We also trained them to use the think-aloud protocol to report their verbal ``walkthroughs'' on thoughts, actions, hypotheses, strategies, goals, and intentions. 
Then we instructed them on how to use the annotation feature to record their answers to the tasks. 
We also told them they could take notes with the annotation feature in the tool if they desired; note-taking on a paper was not permitted.
The participants in DG received an additional instructions on how to use the \tech{} feature. 
The training session lasted about eight minutes on average. 
After the training, we requested they try to solve three practice questions before continuing to the main tasks.
Asking questions was allowed anytime during this tutorial and practice session. 

During the experiment, an experimenter asked participants to ``keep talking'' when they were silent for at least 15 seconds. 
We also requested they report their confidence level when solving each task using a seven-point Likert scale before they proceeded to the next question. 
The experiment took 28 minutes on average ($\sigma=8$ minutes). 
All participants received \$10 for their participation. 
To encourage participants' task performance, we rewarded the top five participants with an additional \$25 as an incentive based on accuracy and task time. 
There was no time restriction, so the participants could have sufficient time for visual exploration. 
We recorded audio and screen activities of the participants for grading verbal walkthroughs, as described in Section~\nameref{sec_coding}.
After the experiment, we conducted a short exit survey in which the participants were asked to answer two questions using a seven-point Likert scale and to write the reasons for their answers. 
The questions were: 1) How much did you like problem-solving? and 2) How much did \tech{} help your analysis? 

\subsection{Equipment}

We used computers with an Intel i7 (3.4GHz) CPU and a 30-inch monitor (2560x1600) throughout the experiment. 
Instead of using a screen-capturing tool that could unexpectedly interrupt the experiment, we wrote a light-weight logger to capture all participant interactions behind.
The logs were categorized using the action taxonomy shown in Figure~\ref{fig_action_taxonomy}. 
we recorded participants' voice and screen activities using high-resolution video cameras for grading verbal walkthroughs.%

\subsection{Measures}

During the experiment, we measured three numeric metrics for evaluation: the task-completion time, accuracy (the number of correct answers divided by five), and the confidence score per task based on the seven-point Likert scale (7 being the most confident).
We logged all participant interactions and captured screen activities along with participant's verbal reports (walkthroughs) through the think-aloud protocol. 
In the next section, we describe how we coded participants' activities from the logs into insight categories and analyzed the logs and video-audio recordings to derive high-level interaction patterns and analysis strategies.

\subsection{Coding Insights and Interactions}
\label{sec_coding}
To obtain a deeper understanding of the problem-solving processes, two of the co-authors of this manuscript individually investigated the think-aloud walkthroughs fromt the recorded video and coded participants' discoveries into five insight categories adapted from the categories proposed in~\cite{Reda15, Guo16}. 
A summary and examples of each insight category are presented in Table~\ref{table_coding_category}.
Note that we were able to analyze participant think-aloud data from 26 participants (9 from CG, 17 from DG), because the other participants did not provide sufficient think-aloud comments for review (e.g., rarely speaking, indistinguishable mumbling). 
Each insight was scored as either 0 or 1 with respect to the five insight categories: Observation, Hypothesis, Question, Goal, and Comment.
For instance, one report could be coded as (0, 0, 0, 0, 1). 
Only one category was assigned to each insight. 
During the investigation, each coder watched the recorded video separately. 
When the participant spoke in the video, each coder determined whether it could be considered as an insight. 
If the coder believed it was an insight, the coder replayed the video to record and categorize a series of interactions according to the interaction taxonomy~\cite{Yi07}. 
Then, we collected the grading results generated by the two coders and compared. 
When there was a disagreement in grading, the coders discussed how to resolve it. 
When the coders did not reach an agreement, the final insight score was produced by calculating an average of the scores. 
The correlation between the grading results and insight types and the scores of the coders was 80.97\%, suggesting the coders were consistent in their grading~\cite{Guo16}.

To extract event types, we analyzed participant logs and recorded videos and observed 11 analysis event types, as shown in Figure~\ref{fig_action_taxonomy}.
On average, each participant performed 10.7 different types of events. 
Each of the event types was categorized into one of four categories: Filter, Reconfigure, Retrieve, and Annotation, as provided in the interaction taxonomy~\cite{Yi07}. 
Note that although annotation is not one of the categories from the taxonomy, but we include it in our work because it was an important activity to summarize intermediate results and to gain insights during the study's analysis session.

\begin{figure}[t]
    \begin{center}
	\includegraphics[width=.95\columnwidth]{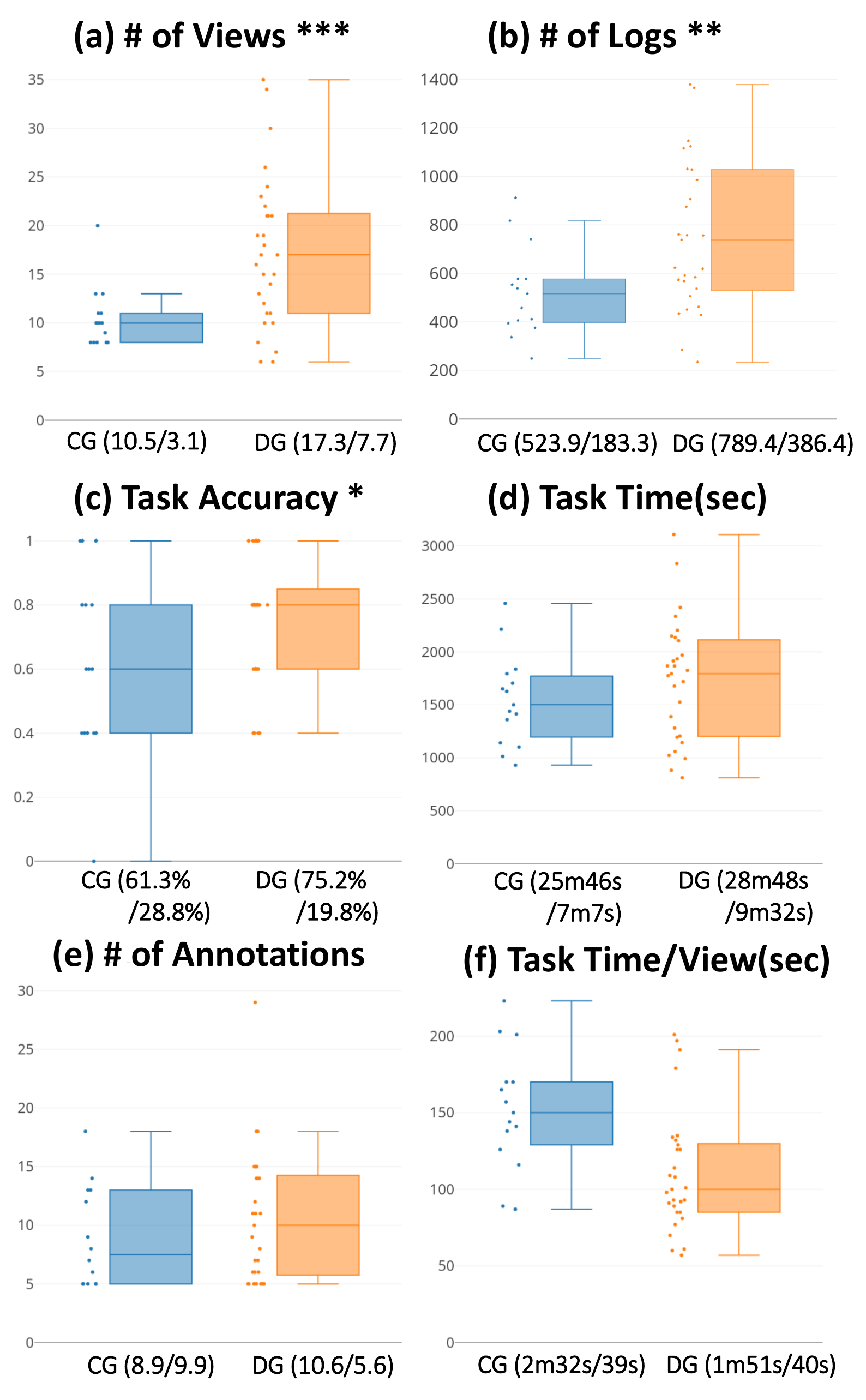}
	\caption{Participants in DG created more views, interactions logs, and more accurately solved problems. Mean and standard deviation values are shown in the parentheses. Each groups' mean and std. pairs are shown in the parentheses. *: $p<$0.05, **:$p<$0.01, ***:$p<$0.001.}
	\label{fig_res_base_dvc}
	\end{center}
\end{figure}

\section{Result Analysis}
\label{sec_result}

In this section, we report the results for use of interactions, task performance, and confidence levels.

\subsection{Users in DG Created More Views and Performed More Interactions}

First, we find in Figure~\ref{fig_res_base_dvc} (a) that  participants who had access to \tech{} ($\mu=17.3$; $\sigma=7.7$) created more views than those in CG ($\mu=10.5$; $\sigma=3.1$), according to Welch's t-test ($t(43)=5.07$, $p<0.001$, Cohen's $d=0.63$ (\textbf{H1 is supported}). 
We conjecture that this effect is probably due to the difference in the perceived interaction costs~\cite{Lam08} in the view creation process between the two groups. 
Creating a new visualization view using the default method requires a sequence of selections, such as deciding the right representation and setting correct coordinates and the parameters, which demands a certain amount of cognitive efforts.
As the number of such sequences increase in a view creation process, participants may experience more burden to perform the correct sequences~\cite{Kobsa01}. 
The burden can be called a ``system-power cost''---the cost to perform system operations~\cite{Lam08}. 
We speculate that the high interaction cost perceived by participants due to the inconvenient view creation process prevents participants from creating more views in CG. 

On the other hand, we believe that \tech{} played an important role in reducing the system-power cost of participants.
Participants described how \tech{} helped their analyses. 
For example, one participant commented \textit{``...I feel it [i.e., \tech{}] helped me keep focusing on my problem-solving process, as it allowed me to easily create other views.''}
Other participants expressed similar opinions--\tech{} allowed participants to create many views by avoiding repeated operations on ``choosing correct views and parameters.''
One participant pointed out that \tech{} removed repeated interactions during visual analysis: \textit{``I used \tech{} a lot, because I did not need to repeat what I should do.''}

\begin{table}[t]
 \scriptsize
 \begin{center}
   \begin{tabular}{cccc}
     \textbf{Control Group (CG)} & \textbf{Interaction Category} & \textbf{Duplication Group (DG)}\\
   \hline
        279.6 ($\sigma=175.9$) & \begin{tabular}[c]{@{}c@{}}\textbf{Retrieve (**)} \end{tabular}  & 527.1 ($\sigma=379.0$)\\ \hline
        97.9 ($\sigma=46.4$) & \begin{tabular}[c]{@{}c@{}}\textbf{Filter (n.s)} \end{tabular} & 131.4 ($\sigma=70.5$)  \\ \hline
        14.5 ($\sigma$=11.6) & \begin{tabular}[c]{@{}c@{}}\textbf{Annotation (n.s)} \end{tabular} & 14.6 ($\sigma=10.1$) \\ \hline
        10.8 ($\sigma=3.2$) & \begin{tabular}[c]{@{}c@{}}\textbf{Create (n.s)}\end{tabular}   & 13.0 ($\sigma=5.0$) \\\hline
        7.1 ($\sigma=7.5$) & \begin{tabular}[c]{@{}c@{}}\textbf{Connect (n.s)} \end{tabular}  & 12.6 ($\sigma=15.1$)\\ \hline
        6.3 ($\sigma=5.8$) & \begin{tabular}[c]{@{}c@{}}\textbf{Reconfigure (**)}  \end{tabular} & 13.6 ($\sigma=10.5$)
   \end{tabular}
    \caption{Performed interactions based on Yi's interaction taxonomy~\cite{Yi07}. 
    }
     \label{tab_avg_action}
 \end{center}
\end{table}

Participants' log data show that participants in DG performed more actions than those in CG ($t(43)=2.94$, $p=0.004$, $d=0.39$, \textbf{H6 is supported}), as Figure~\ref{fig_res_base_dvc}~(b) shows. 
To further analyze participant interactions, we categorize the interactions in Figure~\ref{fig_action_taxonomy} into Yi et al.'s interaction taxonomy~\cite{Yi07}. 
Table~\ref{tab_avg_action} presents our categorization results, where we observe that the participants in DG performed \textit{Retrieve} (e.g., table look-up) and \textit{Reconfigure} (e.g., switching PCP axes) interactions more than those in CG.
Note that we exclude \textit{Select} and \textit{Explore} interactions, as they are not exactly matched to our tool functions (e.g., relocating views, changing view sizes).
Participants reported that they performed many interactions especially when they needed to perform comparisons with many views created by \tech{}.
\textit{``It [\tech{}] was useful, especially when I need to compare visualizations with many views,"} one participant said. %
Another participant also commented that \textit{``I wanted to see results of my brushing in other visualizations for comparison. 
Duplication enabled me to quickly create many other visualization views and perform brushing for the purpose.''}

Participants in DG more frequently used the pin interaction 12.1 times ($\sigma=15.1$) during the experiment, which is significantly higher ($t(24)=2.18$, $p=0.03$, $d=0.27$) than those in CG who used it 4.9 times ($\sigma=4.8$) (\textbf{H7 is supported}).
We believe that using many views could lead to several candidate answers or critical information being pinned ($r=0.37$, $p<0.001$) for reducing temporal-frame association and state-change costs~\cite{Lam12}.
In Figure~\ref{fig_vwh_tree2}, \eg{22} shows an example of a problem solving strategy of a participant who frequently used pins (i.e., yellow strips) on scatterplots to maintain findings in the view and duplicated views to discover more findings.

\begin{table}[t]
 \scriptsize
 \begin{center}
   \begin{tabular}{cccc}
     \textbf{Control Group (CG)} & \textbf{Insight Category} & \textbf{Duplication Group (DG)}\\
   \hline
        11.11($\sigma=3.0$) & \begin{tabular}[c]{@{}c@{}}\textbf{Observation (n.s)} \end{tabular}  & 13.83 ($\sigma=4.87$)\\ \hline
        0.44 ($\sigma=0.68$)& \begin{tabular}[c]{@{}c@{}}\textbf{Hypothesis (**)} \end{tabular} & 2.0 ($\sigma=1.46$)  \\ \hline
        1.11 ($\sigma=1.29$)& \begin{tabular}[c]{@{}c@{}}\textbf{Question (n.s)} \end{tabular}   & 2.47 ($\sigma=3.53$) \\ \hline
        4.67 ($\sigma=2.0$)& \begin{tabular}[c]{@{}c@{}}\textbf{Goal (***)}\end{tabular}   & 8.71 ($\sigma=2.95$) \\\hline
        1.33 ($\sigma=0.94$)& \begin{tabular}[c]{@{}c@{}}\textbf{Comment (*)} \end{tabular}  & 3.88 ($\sigma=3.74$)\\ \hline
        19.0 ($\sigma=4.97$) & \begin{tabular}[c]{@{}c@{}}\textbf{Total (**)}  \end{tabular}  & 31.59 ($\sigma=12.26$)   
   \end{tabular}
    \caption{Scores from grading think-aloud walkthroughs. The DG participants produced more insights than those in the CG. %
    }
     \label{tab_thinkaloud_score}
 \end{center}
\end{table}

\begin{figure}[t]
    \begin{center}
	\includegraphics[width=.99\columnwidth]{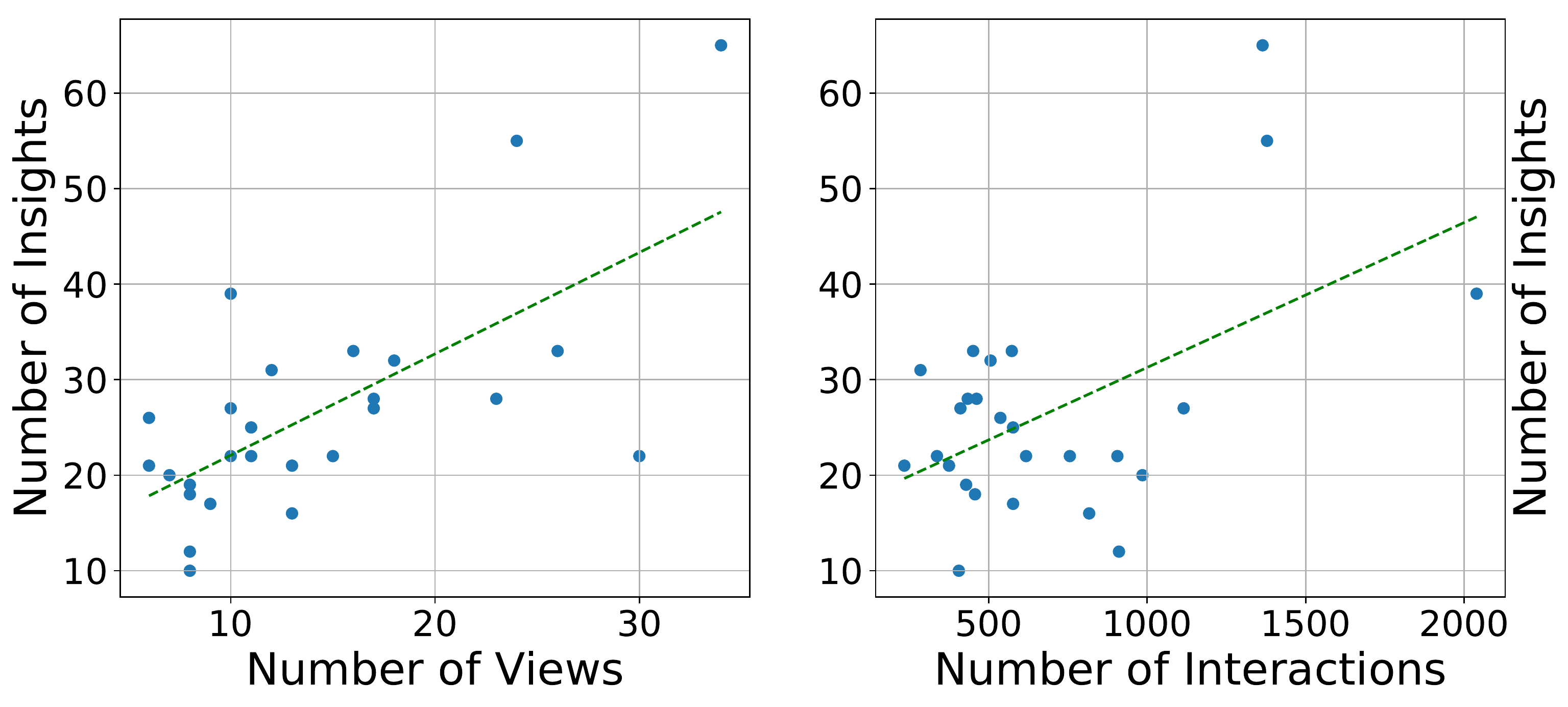}
	\caption{(left) A positive correlation exists between the number of views and number of insights ($r=0.67$, $p<0.001$) and (right) between the number of interactions and number of insights ($r=0.52$, $p=0.006$).	}
	\label{fig_corr_insights}
	\end{center}
\end{figure}

\begin{figure}[t]
    \begin{center}
	\includegraphics[width=.99\columnwidth]{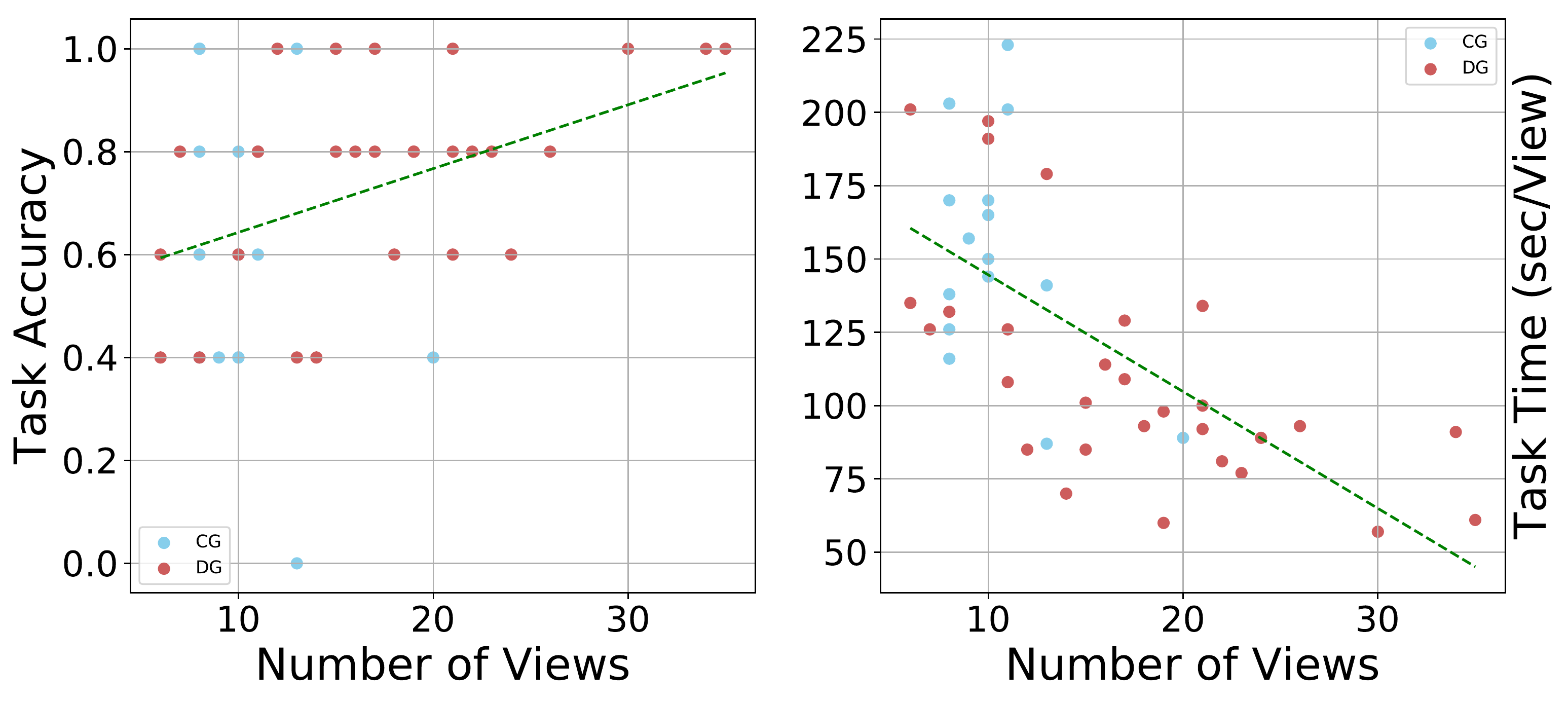}
	\caption{(left) A positive correlation exists between the number of views and task accuracy ($r=0.38$, $p=0.01$), and (right) a negative correlation exists between the number of views and task time per view ($r=-0.66$, $p<0.001$).}
	\label{fig_corr_view}
	\end{center}
\end{figure}

\subsection{Analysis of Insights, Accuracy, and Time}
We analyzed various performance measures based on the quality of analysis, accuracy of results, and analysis time.
Quality of analysis results from graded think-aloud walkthroughs are shown in Table~\ref{tab_thinkaloud_score}.
Participants in DG received higher scores for (1) hypotheses ($t(25)=3.56$, $p=0.002$, $d=1.37$), (2) goals ($t(25)=3.96$, $p<0.001$, $d=1.60$), (3) comments ($t(25)=2.57$, $p=0.018$, $d=0.94$), and (4) the total score ($t(25)=2.84$, $p=0.009$, $d=1.35$) than those in CG (\textbf{H4 is supported}).
In addition, correlation analysis results (Figure~\ref{fig_corr_insights}) indicate that both the number of views ($r=0.67$, $p<0.001$) and interactions ($r=0.52$, $p=0.006$) have positive correlations with the number of generated insights.

Next, we compare accuracy and time for the two groups.
We can see by Figure~\ref{fig_res_base_dvc} (c) that participants in DG received higher scores ($\mu=0.752$; $\sigma=0.20$) than those in CG ($\mu=0.613$; $\sigma=0.29$), which is also supported by Welch's t-test ($t(43)=2.06$, $p=0.04$, $d=0.3$,~\textbf{H2 is supported}); however, there was no difference in the amount of time taken to solve the tasks, as shown in Figure~\ref{fig_res_base_dvc} (d)---\textbf{H3 is rejected}.
Correlation analysis results show a positive correlation (Figure~\ref{fig_corr_view} left) between the number of generated views and accuracy ($r=0.38$, $p=0.01$) and a strong negative correlation (Figure~\ref{fig_corr_view} right) between the number of views and task time per view ($r=-0.66$, $p<0.0001$).

\begin{figure*}[t]
\begin{center}
\includegraphics[width=1.95\columnwidth]{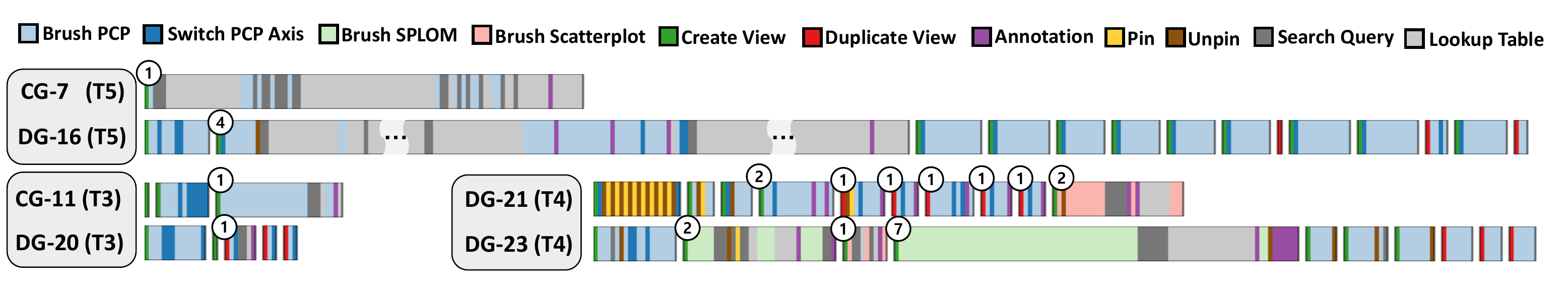}
\caption{Examples of participants' interaction sequences over time. Interactions from different participants are shown using CG and DG labels to indicate experimental group, with task number shown in parentheses. Each horizontal bar represents views used during a task over time, and the colored bands show interactions performed on each view. The circle numbers show the number of annotations created on each view. Participants showed a variety of patterns in the study in terms of the number of views, visualizations used, interactions performed, and annotations.}
\label{fig_vwh_tree}
\end{center}
\end{figure*}

To sum up, participants in DG produced more insights and more accurately completed the given tasks than those without using \tech{} without wasting much time.
There is a significant positive relationship between the number of views and interactions: the more views, the more interactions during analysis (Figure~\ref{fig_corr_insights}, right). %
We also find that the interactions (Figure~\ref{fig_corr_view}, right) and spent time ($r=0.33$, $p=0.028$) for each view are decreased without increase in time for analysis.
We estimate the shift in time allocation is mainly due to the reduced decision costs to create and inspect multiple views, which helped participants use their cognitive efforts to formulate more insights than to make selections for a new view, as is in line with the Lam's framework~\cite{Lam08}.
This result is particularly interesting because it implies that using many views during visual analysis does not increase analysis time significantly.
The results indicate that the number of views is correlated with accuracy; however, the study did not guarantee the causality between them because the participants had the freedom to create as many views as they desired.

\subsection{Confidence Level}

Both groups reported a high confidence level (CG: $\mu=5.9$, $\sigma=0.88$; DG: $\mu=6.0$, $\sigma=0.76$) on the 7-level Likert scale, but there were no significant differences between the two groups related to a confidence level (\textbf{H5 is not supported}). 
Initially, we assumed that participants in DG would have a higher confidence level than those in CG. 
We speculate that participants in both groups showed high confidence in their answers because participants in both groups may have felt that they had enough time to perform the tasks, as reflected in their task time. 
We also think that the participants in the CG might have felt that they could work on the tasks with fewer views due to the perceived difficulty level of the tasks, as described in Sec.~\nameref{sec_subsec_pattern}.
We suspect CG participants were overly confident for less accurate answers, but further investigation would be needed to study such patterns.

\section{Results: Analysis Strategies}
In this section, we report our observations on analysis strategies and annotation utilization patterns (RQ3).
We also provide participant feedback. 

\subsection{Strategies When Many Views Are Available}
\label{sec_subsec_pattern}

\begin{figure}[t]
    \begin{center}
	\includegraphics[width=.99\columnwidth]{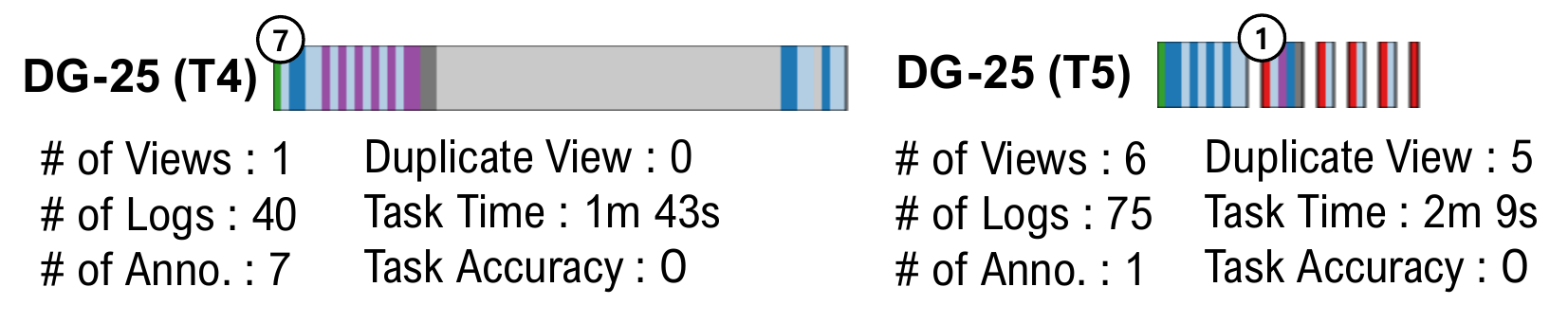}
	\caption{A participant \eg{25} stated that using \tech{} was fun and demonstrated different strategies with task T4 and T5 by intentionally using \tech{}.}
	\label{fig_tree_feedback_participant}
	\end{center}
\end{figure}

We report findings from our analyses of participants' analysis behaviors and strategies, as determined by a qualitative analysis of participants' interaction logs and annotations. 
Our analysis revealed differences in problem-solving patterns and strategies in terms of interaction, view numbers, and visualization. 
We visualize analysis behaviors through Figure~\ref{fig_vwh_tree} to Figure~\ref{fig_vwh_tree2} that show created views over time from left to right.
Colored bands represent interactions and the number of annotations created on a view is shown by circled numbers along the time span.

\textbf{Changing Strategies with Frequent Interactions}:  %
Some participants dramatically changed their dominant interaction patterns from table-lookup to brushing as they began creating many views.
Figure~\ref{fig_vwh_tree} shows an example of two participants: \cg{7} and \eg{16}. 
We observed that \cg{7} frequently used PCP brushing and table lookups to count numbers, as T5 asked participants to find the number of German car models.  %
Participant \eg{16} initially took a similar path: she started with brushing on PCP and frequently performed table lookups. 
Then, \eg{16} soon adopted a different strategy by creating 12 more views. 
As we closely observed the views, we noticed that each view represented PCP, and she created the views to perform brushing. 
As the number of views increased, her dominant interaction also changed from table lookup to PCP brushing, which implies that when many views are available, participants may prefer to quickly switch perspectives (reconfigure) and to brush one view to see changes in another (filter).
This example implies the need for recommendation techniques that can recognize changes in problem-solving strategies and interactions and can recommend efficient interactions with multiple coordinated views during a visual analysis.

\textbf{Easier Context Switching and Branching}:
One expectation of using many views is that participants can be better aided in developing different analysis paths, as participants often encounter roadblocks and must switch to other paths~\cite{Kwon11}.
They sometimes need to take multiple paths in parallel. 
Using multiple views allows users to develop multiple analysis paths.
Participant \cg{11} completed T3 mainly using two PCP views (see Figure~\ref{fig_vwh_tree}). 
In the first PCP view, she mainly performed a series of brushing interactions on the ``Power" dimension. 
Then, she created another PCP view. 
The main interactions for the second view were also brushing, but at times, the interactions focused on the ``Brand" dimension. 
An interesting observation is that she paused for a moment before creating the second view and commented: \textit{``This is not going to end in this way."}
After the pause, she continued her exploration by making another PCP view and performing brushing on another dimension, ``Brand."

Similarly, participant \eg{20} (Figure~\ref{fig_vwh_tree}) used the same series of brushings on the first PCP view and the time of pause; however, interestingly, after the pause, \eg{20} began creating three additional views by using \tech{} and quickly applied a few more brushings for simultaneous comparison. 
While this comparison seems somewhat exaggerated, it implies \tech{}'s potential to not only ease view creation but also to reduce participants’ frustration due to the visualization roadblocks~\cite{Kwon11} during analysis path development. 

Having many views, participants can take alternative paths without losing the current progress and context. 
This pattern of analysis can be considered similar to programmers' branching and merging activities on code repository. 
Programmers can freely test multiple ideas due to the less costly option of ``branching'' rather than directly revising the main version. 
One design implication of this analogy would be related to determining how to help participants ``merge" their different analytic paths. 
Particularly for open-ended investigative analyses, participants might be required to test multiple hypotheses and then to merge them to obtain new insights. 
 
\textbf{Strategies with View Combinations}:
The tasks used for the experiment could be completed using several combinations of visualizations. 
The DG participants utilized various visualization combinations compared to those in the CG. 
In general, the PCP-table combination was a popular combination in which the main strategy is brushing from PCP and lookups in a table, as seen in Figure~\ref{fig_vwh_tree} for participant \cg{7} and Figure~\ref{fig_tree_feedback_participant} for participant \eg{25}. 
The DG participants also tested different visualization combinations, such as PCP-scatterplot and PCP-SPLOM combinations, along with the pin interaction, as shown in Figure~\ref{fig_vwh_tree} for \eg{21} and \eg{23}. 
This may imply that enabling easy view creation can trigger new combinations of system's available visualizations, which could lead to other non-typical but useful visualization combinations and the development of new analysis strategies.

\subsection{Analysis of Annotation Utilization Patterns}
There was no significant difference in the number of created annotations between the two groups. 
Still, qualitatively reviewing the way participants utilized the annotation function during the experiment can help designers develop a useful multi-dimensional data analysis system. 
In this section, we report the annotation observations during the visual analysis. 

\begin{figure}[t]
    \begin{center}
    \includegraphics[width=0.99\columnwidth]{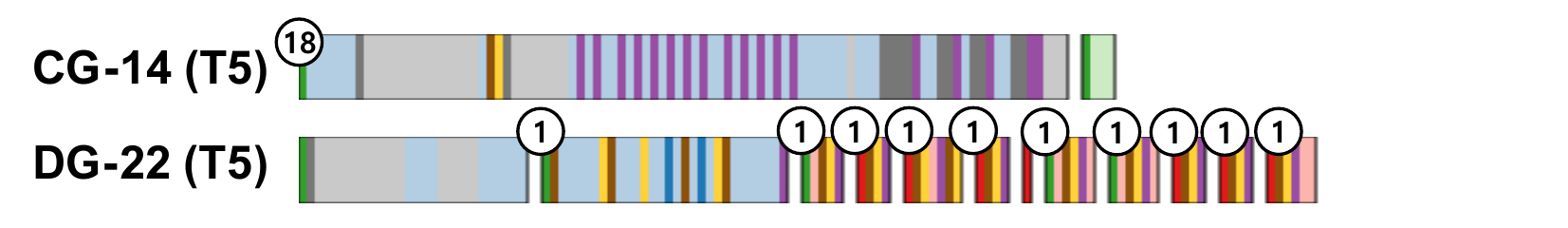}
    \caption{Two observations of problem-solving patterns. \cg{14} created annotations in the first view. \eg{22} frequently pinned visualization states on many views.}
    \label{fig_vwh_tree2}
    \end{center}
\end{figure}

We observed two patterns in terms of the total annotation number. 
The first pattern was only one annotation for a task, which means participants created one annotation to answer each task (e.g., \cg{11} and \eg{20} in Figure~\ref{fig_vwh_tree}). 
Another pattern was creating several annotations. 
We observed that participants created from 7 to 23 annotations in their analyses without counting five annotations left for answers. 
This pattern occurred when participants recognized that they could use annotations as notebooks: \textit{``I can use annotations to leave a brand name on each scatterplot!"} (stated by participant \eg{22} while solving T4 with a scatterplot). 
Later, the annotations used as notebooks were revisited for simultaneous comparisons with calculated and annotated results (i.e., annotations to support limited short-term memory~\cite{Phillips74, Luck97}).
\eg{22}'s work history is presented in Figure~\ref{fig_vwh_tree2}, presenting 10 annotations.

We also observed that the participants left their annotations in three different locations. 
They left their annotations: 1) in the view that was first created (e.g., \cg{14} in Figure~\ref{fig_vwh_tree2}), 2) in the view where they found an answer (e.g., \cg{11} in Figure~\ref{fig_vwh_tree}), or 3) in the view where they performed calculations (e.g., \eg{22} in Figure~\ref{fig_vwh_tree2}).
After analyzing the annotation patterns, we conjectured that participants could have better performed a visual analysis with a visual interface that allows for annotation overviews and organization~\cite{Chin09, Kang09}. 
For example, during the analysis, participant \eg{7} stated: \textit{``I'm going to write down an answer. But which view should I leave my annotation?"} 

We observed participants' interactions for insight management that could be better supported by an annotation organization interface. 
For example, participant \eg{23} (Figure~\ref{fig_vwh_tree}) created her own format for annotation organization and sorting, such as [year, brand, number]. 
Still, we did not observe consistency in participant annotation formats. 
We also assumed that searching, sorting, and filtering functions are useful.
For instance, \eg{7} visited several views to search for a previous annotation. 
Developing a method to distinguish between general and important annotations (e.g., annotation panel) could improve participants' visual analysis with annotations.

\subsection{Participant Feedback}
\label{sec_feedback}

In this section, we report participant feedback. The DG participants rated \tech{} functionality 4.2 out of 7. 
The lower score than the average enjoyment score (5.01) was unexpected because they recorded higher accuracy than those in the CG. 
Based on participant comments, one reason for this low score could be the easy-to-solve tasks. \eg{2}, who selected 4, said, \textit{``... I think I could reduce task time by using \tech{} and performing comparison with many views. 
But the real tasks after the tutorial session looked somewhat simple and could be finished without using \tech{}."} 
Another reason could be that the default view creation option was not cumbersome because a new view could be created within four clicks. \eg{24}, who selected 2, mentioned this: \textit{``I did not see much difference between \tech{} and  `Create' in the creation view."} \eg{24} was one of the participants who did not use \tech{} at all in the DG and recorded low accuracy. 

There were participants who discussed the effectiveness of \tech{}. 
Participant \eg{12} rated the usefulness of \tech{} at 7. 
According to the video analysis, while solving T3 with several table look-ups, \eg{12} said, \textit{``It seems too inefficient to solve in this way."} 
She then used \tech{} to filter the ``brand'' axis in the PCP and said, \textit{``Oh! I can solve it faster by creating four views by using \tech{} now."} 
Participant \eg{16} stated, \textit{``I completely understood the usefulness of \tech{} and enjoyed the tasks by applying different strategies with many views." } 
After reviewing \eg{16}'s visual analysis log and video, we found that she initially solved T3 with one PCP view and table lookups, similar to \cg{7} in Figure~\ref{fig_vwh_tree}. 
Interestingly, \eg{16} used \tech{} often after that point. 
For example, she created 12 more views when completing T4 and T5. 
Participant \eg{16}'s work history is presented in Figure~\ref{fig_vwh_tree}. 
Similarly, \eg{25} changed her problem-solving strategy, as shown in Figure~\ref{fig_tree_feedback_participant}. 
In T4, she initially performed comparisons using PCPs and table lookups, but later, she performed side-by-side comparisons to solve T5. 
These examples demonstrate that using many views could result in participants' analytical strategy changes.

\section{Limitations and Discussion}\label{sec_discussion}
In this section, we discuss the limitations of this work. 
Initially, participants were given few parameters that could be selected for view creation. 
If there were more parameters that could be selected for each view creation (i.e., increased interaction costs), the \tech{} interaction might have been utilized more, and the specifics of interaction behaviors will likely depend on such differences. 
The rationale behind the small number of parameters was that most participants were not familiar with the visualizations and data, and it is often necessary to limit tool complexity for experimentation. 
Performing a study with experts and more complex tasks may also provide additional knowledge of \tech{} or reveal other types of interaction strategies. 
For example, Sariaya et al.~\cite{Saraiya05} called for an approach for the difficulty of selecting many parameters in analyzing bioinformatics data. 
For studying analysis behaviors in future work, it may be useful to observe a more prolonged analysis by experts. 

Also related to limited complexity of the analysis scenario for the purposes of experimentation, the tasks in the study often required simple computation and filtering rather than requiring complex solving methods or inferences. 
For example, one SPLOM visualization may allow users with visualization background to solve T1--T3 quickly. 
Future studies with greater complexity may provide opportunities to study additional interactions and design features, in addition to what various analysis patterns and insights with the tasks and combinations of the visualizations (Figure~\ref{fig_vwh_tree}--Figure~\ref{fig_tree_feedback_participant}).
It is possible that participants may have recorded low accuracy for later questions (e.g., T5), but we intended the order to help participants sufficiently understand the data, before they answer harder questions, as Battle and Heer used in the experiment~\cite{Battle19}. 
We could not observe effective use of the VWH (history view) in the tool, which could be resulted from lack of appropriate functionalities for monitoring, reviewing, and comparing interactions to support what-if scenarios.

\section{Conclusion}
Despite the popularity of CMVs in both the research and application domains, few research studies have focused on the relationship between the number of views and visual analysis results and processes. 
In this work we design a visual tool with visualization duplication that facilitates easy view creation by removing repeated parameter selection when creating views.
The experimental results indicate that users with visualization duplication effectively helps users create a desired number of views. 
The results also reveal that using many views can not only bring better analysis outcomes (i.e., high accuracy, more insights), but also allow various analysis strategies with different interaction sequences.

\begin{acks}
\end{acks}

\bibliographystyle{SageH}
\bibliography{./bibliographies/MCV,./bibliographies/HistoryVis,./bibliographies/DesignModelHCI,./bibliographies/Books,./bibliographies/VisualExploration,./bibliographies/VWMTask,./bibliographies/Insight,./bibliographies/Comparison,./bibliographies/Vistech,./bibliographies/PublicData,./bibliographies/Survey,./bibliographies/ETCBIB}

\end{document}